\documentclass[
floatfix,
twocolumn,
showpacs,showkeys,preprintnumbers,aps,prc]
{revtex4}

\usepackage{amsmath}
\usepackage{amssymb}
\usepackage{revsymb}
\usepackage{graphicx}
\usepackage{dcolumn}

\begin{document}

\title{Covariant calculation of nonstrange decays of strange baryon resonances}
\author{T. Melde, W. Plessas, B. Sengl}
\affiliation{Theoretische Physik, Institut f\"ur Physik, Karl-Franzens-Universit\"at,
Universit\"atsplatz 5, A-8010 Graz, Austria}

\begin{abstract}
We report on a study of $\pi$ and $\eta$ decays of strange baryon resonances 
within relativistic constituent-quark models based on one-gluon-exchange and
Goldstone-boson-exchange dynamics. The investigations are performed in the
point form of Poincar\'e-invariant relativistic quantum mechanics with a 
spectator-model decay operator. The covariant predictions of the
constituent-quark models underestimate the experimental data in most cases.
These findings are congruent with an earlier study of nonstrange baryon decays
in the light-flavor sector. We also consider a nonrelativistic reduction of the 
point-form spectator model, which leads to the decay operator of the
elementary emission model. For some decays the nonrelativistic results differ
substantially from the relativistic ones and they exhibit no uniform behavior
as they scatter above and below the experimental decay widths.
\end{abstract}


\maketitle

\section{Introduction}

Strong decay processes still present a considerable challenge within 
the physics of hadrons. This is unfortunate, not only in view of the 
vast amount of experimental data but also because the decay properties 
of hadron resonances give important insights into strong 
interaction physics (see, for example, the recent NSTAR 
workshops~\cite{Drechsel:2001,Dytman:2003pi,Bocquet:2004}). 
Investigations of strong decay processes date back to the late 
1960s~\cite{Becchi:1966yt,Mitra:1967yt,Faiman:1968js,Faiman:1969at,%
Feynman:1971wr}, and with the refinement of constituent quark models (CQMs)
various aspects of the strong decays have been 
studied~\cite{Koniuk:1980vy,Kumano:1988ga,LeYaouanc:1988aa,%
Stancu:1988gb,Stancu:1989iu,Capstick:1993th,Capstick:1994kb,%
Geiger:1994kr,Ackleh:1996yt,Krassnigg:1999ky,Plessas:1999nb,%
Theussl:2000sj}. In particular, the decay mechanism and 
the type of hyperfine interactions in CQMs have been in the focus of
interest. These investigations have been performed within nonrelativistic
or so-called relativised models, and usually a number of parameters has been
introduced beyond the CQMs employed.
Further complications resulted in the ambiguity
of the proper phase space factor and various forms have been used.
As a consequence the available results are strongly dependent on the chosen
inputs. This makes them hardly comparable to each other, and from the
comparison with experiment the quality of the CQMs cannot be judged reliably.

In our investigations we are primarily interested in the direct predictions
of decay widths by different types of CQMs. Once they are established on
a consistent basis for all decay modes, one can go ahead to study particular
details of the decay mechanism as well as baryon wave functions. 
Recently, we presented a covariant calculation of $\pi$ and $\eta$ decays
of $N$ and $\Delta$ resonances with relativistic CQMs of the one-gluon-exchange
(OGE) and Goldstone-boson-exchange (GBE) types~\cite{Melde:2005hy}.
The investigations were performed in the framework of Poincar\'e-invariant quantum 
mechanics~\cite{Keister:1991sb}. In particular, we adhered to its point-form 
version~\cite{Dirac:1949,Leutwyler:1977vy,Klink:1998pr} and
applied a spectator-model decay operator. In this way the transition amplitude
could be calculated in a manifestly covariant manner and 
ambiguities regarding the phase-space factor could be avoided.

Here we report on the extension of our study of $\pi$ and $\eta$ decays to
strange baryon resonances. Again we work with the relativistic GBE and OGE CQMs
of Refs.~\cite{Glozman:1998ag,Glozman:1998fs} and~\cite{Theussl:2000sj},
respectively.

While there is a wealth of experimental data on these types of decays,
theoretical investigations are rather scarce in the literature, at least within
modern CQMs; in particular, we are not aware of any relativistic calculations.
There exists an older study of strange resonance decays~\cite{Mitra:1967yt} 
but the corresponding results are mainly of a qualitative nature. More recently,
since the advent of CQMs, there have only been a few investigations of strong
decays in the strange 
sector~\cite{Koniuk:1980vy,Capstick:1998uh,Krassnigg:1999ky,Plessas:1999nb}. 

In Section~\ref{sec:theory} we briefly describe the theoretical framework of the
relativistic calculations with the PFSM decay operator; the nonrelativistic limit
is delegated to the Appendix~\ref{sec:appendix}. 
In Sections~\ref{sec:num1} and~\ref{sec:num2} we present the numerical results 
for decay widths in the $\pi$- and $\eta$-channels, respectively.
Finally, in Section~\ref{sec:summary}, we summarize our findings and give
a conclusion.

\section{Theory
\label{sec:theory}}

The decay width of a hadron resonance is defined by the expression
\begin{multline}
\label{eq:decwidth}
{\mit\Gamma}_{i\to f}=\frac{|{\vec q}|}{4M^2}\;\frac{1}{2J+1}
\\ \times
\sum\limits_{M_J,M_{J'}}
\frac{1}{2T+1}\sum\limits_{\;M_T,M_{T'},M_{T_m}}
|F_{i\to f}|^2,
\end{multline}
with the transition amplitude $F_{i\to f}$ given by the matrix element
of the four-momentum conserving reduced decay operator $\hat D^m_{rd}$
between incoming and outgoing hadron states
\begin{multline}
\label{eq:transel}
F_{i\to f}=\\
\langle V',M',J',M_{J'},T',M_{T'}|{\hat D}^m_{rd}
|V,M,J,M_J,T,M_T \rangle \, ,
\end{multline}
where $m$ refers to the particular mesonic decay mode.
In our case, $q_\mu=(q_0,\vec q)$ denotes the four-momentum of the outgoing
meson in the rest-frame of the decaying baryon resonance. The latter is
expressed by the eigenstate $|V,M,J,M_J,T,M_T \rangle$,
characterized by the eigenvalues of the
velocity $V$, mass $M$, intrinsic spin $J$ with z-component $M_J$, and isospin
$T$ with z-projection $M_T$; correspondingly the outgoing baryon state is
denoted by primed eigenvalues. The baryon eigenstates are obtained by the
solution of the eigenvalue problem of the invariant mass operator $\hat M$
including the interactions. They are simultaneously eigenstates of the baryon
four-velocity $\hat V^\mu$.

Representing the baryon eigenstates with suitable basis states the matrix
element in Eq.~(\ref{eq:transel}) can be evaluated by the following integral
\begin{widetext}%
\begin{eqnarray}
&&
    {\langle V',M',J',M_{J'},T',M_{T'}|{\hat D}_{rd}^m|
    V,M,J,M_J,T,M_T\rangle}
    =
     {\frac{2}{MM'}\sum_{\sigma_i\sigma'_i}\sum_{\mu_i\mu'_i}{
	\int{
	d^3{\vec k}_2d^3{\vec k}_3d^3{\vec k}'_2d^3{\vec k}'_3
	}} } 
\nonumber\\
&&
{\times \sqrt{\frac{\left(\sum_i \omega'_i\right)^3}
	{\prod_i 2\omega'_i}}
	\Psi^\star_{M'J'M_{J'}T'M_{T'}}\left({\vec k}'_1,{\vec k}'_2,{\vec k}'_3;
	\mu'_1,\mu'_2,\mu'_3\right)
	\prod_{\sigma'_i}{D_{\sigma'_i\mu'_i}^{\star \frac{1}{2}}
	\left\{R_W\left[k'_i;B\left(V'\right)\right]\right\}
	}
	} 
\nonumber\\
&&
     {\times
     \left<p'_1,p'_2,p'_3;\sigma'_1,\sigma'_2,\sigma'_3\right|{\hat D}_{rd}^m
	\left|p_1,p_2,p_3;\sigma_1,\sigma_2,\sigma_3\right>
	\nonumber } 
\nonumber\\
&&
	\times \prod_{\sigma_i}{D_{\sigma_i\mu_i}^{\frac{1}{2}}
	\left\{R_W\left[k_i;B\left(V\right)\right]\right\}}
	     {\sqrt{\frac{\left(\sum_i \omega_i\right)^3}
	{\prod_i 2\omega_i}}
	\Psi_{MJM_J TM_T}\left({\vec k}_1,{\vec k}_2,{\vec k}_3;\mu_1,
	\mu_2,\mu_3\right)} \, .
\end{eqnarray}
\end{widetext}%
Herein,
$\Psi_{MJM_J TM_T}\left({\vec k}_1,{\vec k}_2,{\vec k}_3;\mu_1,
	\mu_2,\mu_3\right)$ represents the rest-frame wave function of the
incoming baryon, where the $\mu_i$ refer to the spin projections of the
three quarks $i=1,2,3$, and their three-momenta $\vec k_i$ sum up to zero; 
analogously,
$\Psi^\star_{M'J'M_{J'}T'M_{T'}}\left({\vec k}'_1,{\vec k}'_2,{\vec k}'_3;
\mu'_1,\mu'_2,\mu'_3\right)$ is the rest-frame wave function of the
outgoing baryon. These wave functions result from the velocity-state
representations of the baryon eigenstates (for more details of the
formalism see Ref.~\cite{Melde:2005hy}). The Wigner rotations stem
from the boosts of the baryon eigenstates relating the individual quark
momenta through $p_i=B(V)k_i$.
The momentum representation of the decay operator follows from the PFSM
construction, which assumes that only one of the quarks directly couples
to the emitted meson, while the other two act as spectators:
\begin{multline}
\label{momrepresent}
\langle p'_1,p'_2,p'_3;\sigma'_1,\sigma'_2,\sigma'_3
|{\hat D}^{m}_{rd}|
 p_1,p_2,p_3;\sigma_1,\sigma_2,\sigma_3\rangle
 \\
 =-3{\mathcal{N}}\frac{ig_{qqm}}{2m_1}\frac{1}{\sqrt{2\pi}}
\bar{u}(p_1',\sigma_1')\gamma_5\gamma^\mu \mathcal{F}^m 
 u(p_1,\sigma_1)q_\mu\\
 2p_{20}\delta^3\left(
{\vec {p}}_2-{\vec {p}}'_2\right)\delta_{\sigma_2 \sigma'_2}
2p_{30}\delta^3\left({\vec {p}}_3-{\vec {p}}'_3\right)
\delta_{\sigma_3 \sigma'_3} \, .
\end{multline}
Here, $g_{qqm}$ is the quark-meson coupling constant, $m_1$
the mass of the active quark, $\mathcal{F}^m$
the flavor-transition operator specifying the particular decay mode,
and $u(p_1,\sigma_1)$ the quark spinor according to the
standard notation~\cite{Yndurain:1996aa}; details of the formalism can
be found in Ref.~\cite{Sengl:2006}.

The factor $\mathcal{N}$ is specific for the PFSM
construction~\cite{Melde:2004qu} and is taken to be
\begin{equation}
{\cal N}=
\left(\frac{M}{\sum_i{\omega_i}}
\frac{M'}{\sum_i{\omega'_i}}\right)^{\frac{3}{2}}.
\end{equation}
This form is congruent with the calculations in Ref.~\cite{Melde:2005hy}
and also consistent with the requirements of baryon charge normalisation
as well as time-reversal invariance of the electromagnetic
form-factors~\cite{Wagenbrunn:2005wk}. The same normalisation factor
was also used in previous studies of the electroweak structure of the
nucleons and the other light and strange baryon ground
states~\cite{Wagenbrunn:2000es,Glozman:2001zc,Boffi:2001zb,Berger:2004yi}. 

We also consider a nonrelativistic reduction of the PFSM decay operator.
Its derivation is given in the Appendix~\ref{sec:appendix} and it leads to the
standard impulse approximation according to the elementary emission model (EEM):
\begin{multline}%
\label{galiinv}
\langle p'_1,p'_2,p'_3;\mu'_1,\mu'_2,\mu'_3
|{\hat D}^{m,NR}_{rd}|
 p_1,p_2,p_3;\mu_1,\mu_2,\mu_3\rangle
 \\
 \propto \frac{\mathcal{F}^m}{2m_1}
\left\{{\vec \sigma}_1 
 \cdot {\vec{q}}-\frac{\omega_m}{2m_1}
{\vec \sigma}_1\cdot\left({\vec {p}}_1+{\vec {p}}'_1\right)
\right\}
\\
\times 2p_{20}\delta^3\left(
{\vec {p}}_2-{\vec {p}}'_2\right)\delta_{\mu_2 \mu'_2}
2p_{30}\delta^3\left({\vec {p}}_3-{\vec {p}}'_3\right)
\delta_{\mu_3 \mu'_3} \, .
\end{multline}%

\section{Constituent-Quark Models
\label{sec:CQMs}}

\begin{figure*}
\includegraphics[height=7.5cm]{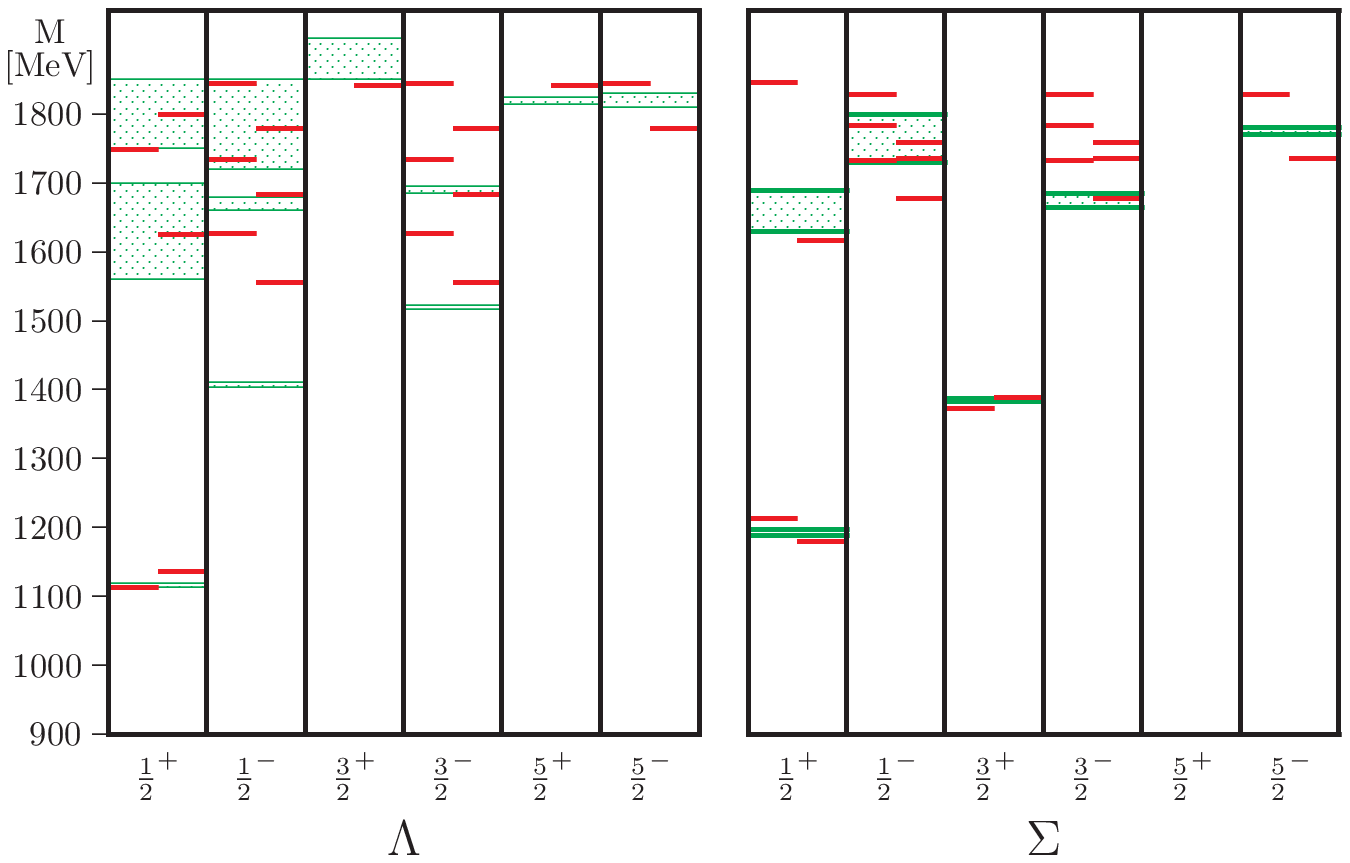}
\hspace{0.05cm}
\includegraphics[height=7.5cm]{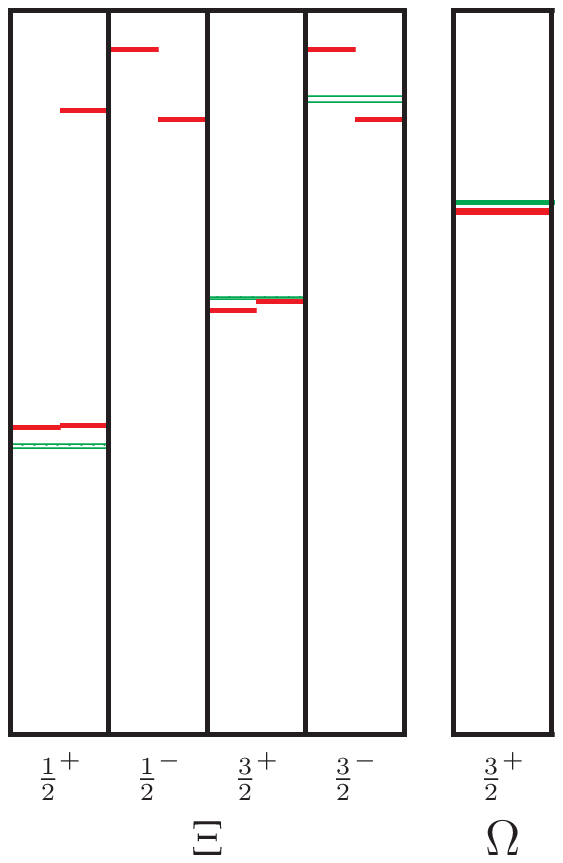}
\caption{Energy levels (solid lines) of the lowest $\Lambda$, $\Sigma$, $\Xi$, 
and $\Omega$ states with intrinsic spin and parity $J^P$ for the OGE
(left levels) and GBE (right levels) CQMs as presented in Refs.~\cite{Theussl:2000sj} 
and~\cite{Glozman:1998ag}, respectively. The shadowed boxes represent the
experimental values with their uncertainties~\cite{PDBook}.
\label{fig:LaSig}
}
\end{figure*}

For the calculations of the mesonic decay widths we employ two different
kinds of relativistic CQMs. Thereby we can learn about the importance of
distinct hyperfine interactions. In particular, we consider the GBE
CQM~\cite{Glozman:1998ag} and the OGE CQM in the variant of the relativistic
version of the Bhaduri-Cohler-Nogami model~\cite{Bhaduri:1981pn} as
parameterised by Theussl et al.~\cite{Theussl:2000sj}. The invariant
mass spectra of both CQMs are shown in Fig.~\ref{fig:LaSig} in comparison
to experiment as compiled by the Particle Data Group (PDG)~\cite{PDBook}. 

The spectra of the two CQMs show the typical behavior as it is well known
from the literature~\cite{Glozman:1998fs}. Only the flavor-dependent
hyperfine interaction of the GBE CQM is able to reproduce at the same
time the correct level orderings in the nucleon and the $\Lambda$
excitation spectra~\cite{Glozman:1998ag}. Both types of CQMs fail in
reproducing the $\Lambda$(1405) resonance. Further shortcomings of the
CQMs may also reside in other strange baryon excitations, for which no
experimental data exist at the moment. Differences
between theoretical and experimental resonance masses, however, can have
a strong influence on the predictions for decay widths. In order to have
these mass effects under control, we calculated the decay widths by using
also experimental masses as input. In this way we can directly
estimate the effects from different quark-model wave functions too.  

\section{Direct Predictions of $\pi$ Decay Widths
\label{sec:num1}}

\renewcommand{\arraystretch}{1.5}
\begin{table*}
\caption{Covariant predictions for $\pi$ decay widths by the GBE
CQM~\cite{Glozman:1998ag} and the OGE CQM~\cite{Theussl:2000sj}
in comparison to experiment. 
The first three columns classify the decaying resonance according
to the PDG~\cite{PDBook}. The relativistic calculations have been
performed along the PFSM, and the EEM results represent their nonrelativistic
limits. We used both theoretical and experimental masses (best estimates of
the PDG) as input. For comparison we present in the last column also the
results of a nonrelativistic calculation by Koniuk and Isgur~\cite{Koniuk:1980vy}.
\label{tab1}
}
\begin{ruledtabular}
{\begin{tabular}{@{}lcr| cccc| cccc| c@{}}
&&
&\multicolumn{4}{c}{Theoretical Mass}
&\multicolumn{4}{c}{Experimental Mass}
&\multicolumn{1}{c}{}\\
Decay &$J^P$&Experiment \small [MeV]
&\multicolumn{2}{c}{Relativistic}
&\multicolumn{2}{c|}{Nonrel. EEM}
&\multicolumn{2}{c}{Relativistic}
&\multicolumn{2}{c|}{Nonrel. EEM}
&\multicolumn{1}{c}{Literature}
\\
\small $\rightarrow \Sigma\pi $&&
& GBE & OGE & GBE & OGE  
& GBE & OGE & GBE & OGE 
& KI \\
\hline
$\Lambda(1405)$&
$\frac{1}{2}^-$&
$\left(50\pm2\right)$ 
    &$55$ 
    &$78$ 
    &$320$ 
    &$611$ 
    &$15$ 
    &$17$ 
    &$76$ 
    &$112$ 
    &$55$
\\ 
$\Lambda(1520)$&
$\frac{3}{2}^-$&
$\left(6.55\pm0.16\right)_{-0.04}^{+0.04}$ 
    &$5$ 
    &$9$ 
    &$5$ 
    &$8$ 
    &$2.8$ 
    &$3.1$ 
    &$2.1$ 
    &$2.3$ 
    &$7.8$
\\ 
$\Lambda(1600)$&
$\frac{1}{2}^+$&
$\left(53\pm38\right)_{-10}^{+60}$ 
    &$3$ 
    &$33$ 
    &$2$ 
    &$34$ 
    &$3$ 
    &$17$ 
    &$1.2$ 
    &$15$ 
    &$14$
\\ 
$\Lambda(1670)$&
$\frac{1}{2}^-$&
$\left(14.0\pm5.3\right)_{-2.5}^{+8.3}$ 
    &$69$ 
    &$103$ 
    &$620$ 
    &$1272$ 
    &$68$ 
    &$94$ 
    &$572$ 
    &$1071$ 
    &$10$
\\ 
$\Lambda(1690)$&
$\frac{3}{2}^-$&
$\left(18\pm6\right)_{-2}^{+4}$ 
    &$19$ 
    &$25$ 
    &$24$ 
    &$28$ 
    &$18$ 
    &$21$ 
    &$23$ 
    &$22$ 
    &$44$
\\ 
$\Lambda(1800)$&
$\frac{1}{2}^-$&
$seen$ 
    &$68$ 
    &$101$ 
    &$473$ 
    &$1175$ 
    &$70$ 
    &$95$ 
    &$485$ 
    &$1095$
    &$121$
\\ 
$\Lambda(1810)$&
$\frac{1}{2}^+$&
$\left(38\pm23\right)_{-10}^{+40}$  
    &$3.8$ 
    &$2.1$ 
    &$55$ 
    &$150$ 
    &$4.1$ 
    &$5.0$ 
    &$55$ 
    &$94$ 
    &$36$
\\ 
$\Lambda(1830)$&
$\frac{5}{2}^-$&
$\left(52\pm19\right)_{-12}^{+11}$ 
    &$14$ 
    &$19$ 
    &$16$ 
    &$24$ 
    &$16$ 
    &$20$ 
    &$22$ 
    &$24$ 
    &$59$
\\ 
\hline 
$\Sigma(1385)$&
$\frac{3}{2}^+$&
$\left(4.2\pm0.5\right)_{-0.5}^{+0.7}$ 
    &$3.1$ 
    &$0.5$ 
    &$6.5$ 
    &$1.1$ 
    &$2.0$ 
    &$2.1$ 
    &$4.1$ 
    &$4.8$ 
    &$7.8$
\\ 
$\Sigma(1660)$&
$\frac{1}{2}^+$&
$seen$ 
    &$10$ 
    &$24$ 
    &$2$ 
    &$15$ 
    &$12$ 
    &$14$ 
    &$2.4$ 
    &$6.9$ 
    &$14$
\\ 
$\Sigma(1670)$&
$\frac{3}{2}^-$&
$\left(27\pm9\right)_{-6}^{+12}$ 
    &$15$ 
    &$23$ 
    &$21$ 
    &$32$ 
    &$13$ 
    &$17$ 
    &$17$ 
    &$21$
    &$44$
\\ 
$\Sigma(1750)^1$&
$\frac{1}{2}^-$&
$\left(3.6\pm3.6\right)_{-0}^{+5.6}$ 
    &$58$ 
    &$102$ 
    &$480$ 
    &$1249$ 
     &$63$ 
    &$102$ 
    &$574$ 
    &$1402$ 
    &$$
    \\ 
$\Sigma(1750)^2$&
$\frac{1}{2}^-$&
$\left(3.6\pm3.6\right)_{-0}^{+5.6}$ 
    &$32$ 
    &$44$ 
    &$135$ 
    &$312$ 
    &$32$ 
    &$38$ 
    &$136$ 
    &$262$ 
    &$$
\\ 
$\Sigma(1750)^3$&
$\frac{1}{2}^-$&
$\left(3.6\pm3.6\right)_{-0}^{+5.6}$ 
    &$10$ 
    &$1.0$ 
    &$116$ 
    &$34$ 
    &$10$ 
    &$0.9$ 
    &$110$ 
    &$32$ 
    &$0.25$
\\ 
$\Sigma(1775)$&
$\frac{5}{2}^-$&
$\left(4.2\pm1.8\right)_{-0.3}^{+0.8}$ 
    &$1.9$ 
    &$3.8$ 
    &$2.9$ 
    &$6.9$ 
    &$2.2$ 
    &$3.2$ 
    &$3.5$ 
    &$5.3$ 
    &$6$
\\ 
$\Sigma(1940)$&
$\frac{3}{2}^-$&
$seen$ 
    &$2.2$ 
    &$3.7$ 
    &$0.5$ 
    &$1.1$ 
    &$4.9$ 
    &$5.8$ 
    &$1.6$ 
    &$2.4$ 
    &$19$
\\ 
\hline
\multicolumn{1}{c}{\small $\rightarrow \Lambda\pi $}&
\multicolumn{6}{l}{}\\
\hline
$\Sigma(1385)$
&
$\frac{3}{2}^+$&
$\left(31.3\pm 0.5\right)_{-4.3}^{+4.4}$ &
$11$ &
$11$ & 
$25$ &
$28$ &
$14$ &
$13$ & 
$31$ &
$32$ &
$44$ \\ 
$\Sigma(1660)$
&
$\frac{1}{2}^+$&
$seen$ &
$8$ &
$5$&  
$6$ &
$0.02$ &
$10$ &
$3$&  
$8$ &
$0.05$&
$8.4$\\ 
$\Sigma(1670)$
&
$\frac{3}{2}^-$&
$\left(6\pm3\right)_{-1}^{+3}$ &
$2.5$ &
$2.0$& 
$5.5$ &
$5.1$&
$2.7$ &
$1.5$& 
$6.0$ &
$3.2$ &
$5.8$ \\ 
$\Sigma(1750)^1$
&
$\frac{1}{2}^-$&
$seen$ &
$1.6$ &
$1.5$& 
$43$ &
$67$ &
$0.8$ &
$1.4$& 
$49$ &
$70$ &
$$\\ 
$\Sigma(1750)^2$
&
$\frac{1}{2}^-$&
$seen$ &
$19$ &
$25$& 
$160$ &
$422$&
$18$ &
$25$ & 
$169$ &
$359$  &
$$\\ 
$\Sigma(1750)^3$
&
$\frac{1}{2}^-$&
$seen$ &
$1.0$ &
$2.8$& 
$18$ &
$105$ &
$0.9$ &
$3$& 
$18$ &
$97$&
$28$\\ 
$\Sigma(1775)$
&
$\frac{5}{2}^-$&
$\left(20\pm 4\right)_{-2}^{+3}$ &
$6$ &
$10$& 
$10$ &
$21$ &
$8$ &
$8$& 
$15$ &
$15$ &
$22$\\ 
$\Sigma(1940)$
&
$\frac{3}{2}^-$&
$seen$ &
$0.2$ &
$0.4$ & 
$1.7$ &
$3.5$ &
$0.5$ &
$0.5$ & 
$5.9$ &
$6.1$ &
$0.16$\\ 
\hline
\multicolumn{1}{c}{\small $\rightarrow \Xi\pi $}&
\multicolumn{6}{l}{}\\
\hline
$\Xi(1530)$&
$\frac{3}{2}^+$&
$\left(9.9\right)_{-1.9}^{+1.7}$ &
$2.2$ &
$1.3$&
$4.4$ &
$3.0$&
$5.5$ &
$5.3$&
$11.4$ &
$12.5$ &
$$ \\ 
$\Xi(1820)$
&
$\frac{3}{2}^-$&
$seen$ &
$0.4$ &
$1.6$&
$0.3$ &
$1.4$&
$0.7$ &
$1.2$&
$0.6$ &
$0.9$&
$$ \\
\end{tabular}}
\end{ruledtabular}
\end{table*}

In Table~\ref{tab1} we present the direct predictions of the $\pi$ decay widths
for strange baryon resonances. The relativistic results have been obtained with
the PFSM decay operator with pseudovector coupling as specified in
Eq.~(\ref{momrepresent}). The
nonrelativistic results correspond to the calculation with the EEM transition
operator as given in Eq.~(\ref{galiinv}). For both cases theoretical masses (as
predicted by the particular CQMs) and experimental masses (as quoted by
the PDG~\cite{PDBook}) have been employed.

In general the present results for the $\pi$ decay widths of the strange baryon
resonances parallel the ones obtained earlier in case of the nonstrange
resonances~\cite{Melde:2005hy}: the covariant predictions
usually underestimate the experimental data or at most reach them from below.
Here, there are only a few notable exceptions, namely the decays of $\Lambda$(1405)
and $\Lambda$(1670) going to $\Sigma\pi$. In the first case the overshooting
of the experimental value is only present, if the theoretical resonance masses
are used. It disappears when employing the experimental resonance mass. Therefore
we may attribute the large values for the decay widths essentially to the
theoretical overpredictions of the $\Lambda$(1405) mass, both by the GBE
and OGE CQMs. The situation is not so clear-cut with regard to the
$\Lambda$(1670). Its resonance mass is more or less reproduced in accordance
with the experimental data, at least in case of the GBE CQM; still the
decay widths are predicted far too high. There is only a minor mass effect
in these overpredictions, since they are also not reduced when employing
the experimental resonance mass. Therefore we may suspect the large $\pi$ decay
widths of $\Lambda$(1670) to be caused by another reason, possibly
a coupling of resonance states.

Of particular interest is the decay of the $\Sigma$(1750) $\frac{1}{2}^-$
resonance to $\Sigma\pi$. From the spectrum as presented in Fig.~\ref{fig:LaSig}
we observe three theoretical levels for each CQM, and it appears natural to
identify the lowest lying $J^P=\frac{1}{2}^-$ state with the $\Sigma$(1750)
resonance, which is the only $\frac{1}{2}^-$ $\Sigma$ excitation with at least
three-star status in the PDG compilation.
The predictions for the $\pi$ decay widths of this lowest
lying state turn out to be much bigger than the experimental value measured 
for the $\Sigma$(1750); the corresponding figures can be found under the
entry of $\Sigma$(1750)$^1$ in Table~\ref{tab1}. However, we may also consider
the two other theoretical levels in the $J^P=\frac{1}{2}^-$ excitation
spectrum as candidates for $\Sigma$(1750). Upon calculating their $\pi$
decay widths we find the predictions as given under the entries of
$\Sigma$(1750)$^2$ and $\Sigma$(1750)$^3$ in Table~\ref{tab1}. Surprisingly,
the theoretical results for the last one are pretty consistent with the
magnitude of the experimental value for $\Sigma$(1750). It is thus
suggested to identify the third level $\Sigma$(1750)$^3$ with the
experimentally measured $\Sigma$(1750). The two remaining eigenstates
are then left to be interpreted as the lower lying resonances $\Sigma$(1620)
and maybe $\Sigma$(1560), which are observed in experiment with only two-star
status~\footnote{We note that the eigenstates $\Sigma$(1750)$^2$ and
$\Sigma$(1750)$^3$ can be distinguished by their internal spin structure.
It turns out that their ordering is reversed in the OGE CQM, namely the
$\Sigma$(1750)$^3$ falls below the $\Sigma$(1750)$^2$.}.

The influences from different hyperfine interactions in the CQMs can be
estimated by comparing the results obtained with the experimental masses
in the eighth and ninth columns of Table~\ref{tab1}. In general, they
are small. Considerable differences
are seen only for the $\Lambda$(1600) and $\Sigma$(1750)$^3$ in the
$\Sigma \pi$ channel as well as for $\Sigma$(1660) and $\Sigma$(1750)$^3$
in the $\Lambda \pi$ channel. 

Let us finally have a look at the results from the nonrelativistic
reduction of the PFSM, leading to the EEM. One can hardly find a common
trend among the nonrelativistic EEM predictions.
Rather they scatter below and above the
experimental data. Evidently, the nonrelativistic approximation causes
huge enhancements of the decay widths for the $J^P=\frac{1}{2}^-$
resonances. They become way too high as compared to experiment. On the
other hand, the $J^P=\frac{1}{2}^+$ decay widths are much reduced by the
nonrelativistic approximation, with the exception of the
$\Lambda$(1810). For the $J^P=\frac{3}{2}^-$ resonances
the nonrelativistic results are very similar to the relativistic ones,
with the exception of the $\Sigma$(1940).
As we have kept the phase-space factor fixed, this behavior of the 
nonrelativistic approximation is governed only by the
truncation in the spin couplings and the elimination of Lorentz boosts. 
This makes the effects of the nonrelativistic reduction strongly 
dependent on the decaying resonance. As a result any nonrelativistic
approximation for calculating decay widths must be taken with considerable
doubt.

In the last column of Table~\ref{tab1} we also
quote the predictions of Koniuk and Isgur (KI)~\cite{Koniuk:1980vy} and we
observe a completely different behaviour than in any of our calculations. 
Especially, the KI results are seen in rather good agreement with experimental
data (except for the case of $\Lambda$(1690), whose decay width comes out too
large). It has to be noted, however, that KI introduced additional
parameters to fit their quark model predictions to experiment in order
to generally investigate the feasibility of decay calculations within
constituent quark models. We, on the other hand, refrained from applying
any parameterisation beyond the direct PFSM predictions quoted in
Table~\ref{tab1}, as we are
interested in the pure nature of the relativistic results and their
dependences on the dynamics of different CQMs. Once this step is
clarified, we can proceed to refine the decay calculations in order to
possibly arrive at a more convincing description of hadronic
decays~\footnote{E.g., one could simply start with varying the quark-meson
coupling constant in the decay operator of Eq.~(\ref{momrepresent});
so far we adhered to an $SU(3)$ symmetric choice just as it is employed
in the GBE CQM~\cite{Glozman:1998ag,Glozman:1998fs}. Of course, one should
consider also more substantial improvements of the decay mechanism and
in addition of the description of resonance states.}.
\section{Direct Predictions of $\eta$ Decay Widths 
\label{sec:num2}}
As a second nonstrange decay mode of strange baryon resonances we consider
the $\eta$ decays of $\Lambda$ and $\Sigma$, the results of which are given
in Table~\ref{tab2}. Regarding the predictions of the GBE CQM in case of
theoretical masses we observe that the $\eta$ decays of the $\frac{1}{2}^-$
$\Lambda$(1670) and $\frac{3}{2}^-$ $\Lambda$(1690) resonances (which are
degenerate in our calculation) are not possible energetically, since the
mass eigenvalue of 1136 MeV for the $\Lambda$ ground state lies slightly
too high. On the other hand, the $\Lambda$(1670) decay width of the OGE
CQM obviously results too big, in accordance with the fact that the
$\Lambda$(1670) mass lies too high. These deficiencies become repaired
when the experimental masses are employed. The $\eta$ decay widths of
the $\Lambda$(1670) then come out reasonably for both the GBE and OGE CQMs,
and they are found in agreement with the experimental data as well as the
KI results. The $\eta$ decay width of the $\Lambda$(1690) is in all instances
extremely small.

The biggest $\eta$ decay width is predicted for the $\frac{1}{2}^-$
$\Lambda$(1800) resonance by both the GBE and OGE CQMs; also the
corresponding KI result is the largest one among all $\eta$ decays.
Already the $\pi$ decay width of this state has been found to be rather
large in all cases above (cf., Table~\ref{tab1}). The PDG does not
present any data for nonstrange partial decay widths. Given the fact
that the total decay width of $\Lambda$(1800) is of the order of 200--400
MeV~\cite{PDBook}, the interpretation of the large $\pi$ and $\eta$ decays
should not pose any particular problem, however.  

The remaining $\eta$ decays of the $\Lambda$(1810) and $\Lambda$(1830)
are predicted to be rather small by both the GBE and OGE CQMs. Similar
results are reported also from the KI calculation.

Regarding the $\eta$ decays of the $\Sigma$(1750) resonance we again quote
the widths of all three states that can a-priori be considered as candidates for
this resonance. The relatively largest decay width is obtained by the GBE CQM
for the $\Sigma(1750)^3$ state. It almost reaches the experimental data band
from below. In the previous section this state was interpreted as the proper
candidate for the $\Sigma$(1750) resonance, whereas the $\Sigma(1750)^1$ and
the $\Sigma(1750)^2$ states should rather be identified with the $\Sigma(1560)$
and $\Sigma(1620)$ resonances, respectively. We now find this interpretation
further substantiated by the $\eta$ decays.

In summary we note that the only $\eta$ decays with appreciable decay widths
are the ones of $\Lambda$(1670), $\Lambda$(1800), $\Lambda$(1830), and
$\Sigma$(1750). It is interesting to note that the octet partners of the
former two in the light-flavor sector, namely, the
$N$(1535) and $N$(1650) resonances, are just the ones with
appreciable sizes for $N \to N\eta$ decay widths.

\renewcommand{\arraystretch}{1.5}
\begin{table*}
\caption{Same as Table~\ref{tab1}, but for the $\eta$-decay channels.
\label{tab2}
}
\begin{ruledtabular}
{\begin{tabular}{@{}lcr| cccc| cccc| c@{}}
&&
&\multicolumn{4}{c}{Theoretical Mass}
&\multicolumn{4}{c}{Experimental Mass}
&\multicolumn{1}{c}{}\\
Decay &$J^P$&Experiment \small [MeV]
&\multicolumn{2}{c}{Relativistic}
&\multicolumn{2}{c}{Nonrel. EEM}
&\multicolumn{2}{c}{Relativistic}
&\multicolumn{2}{c}{Nonrel. EEM}
&\multicolumn{1}{c}{Literature}
\\
\small $\rightarrow \Lambda\eta $&&
& GBE & OGE & GBE & OGE  
& GBE & OGE & GBE & OGE 
& KI \\
\hline
$\Lambda(1670)$&
$\frac{1}{2}^-$&
$\left(6.1\pm2.6\right)_{-2.5}^{+3.8}$ 
         & $-$ 
         & $19$ 
         & $-$ 
         & $151$
         & $4$ 
         & $6$ 
         & $22$ 
         & $44$
         & $4.8$
\\
$\Lambda(1690)$
&$\frac{3}{2}^-$
&
         & $-$ 
         & $0.2$ 
         & $-$ 
         & $0.08$
         & $0.02$ 
         & $0.02$ 
         & $\approx 0$ 
         & $\approx 0$
         & $0.01$
\\
$\Lambda(1800)$
&$\frac{1}{2}^-$
&
         & $43$ 
         & $65$ 
         & $223$ 
         & $624$
         & $46$ 
         & $62$ 
         & $264$ 
         & $526$
         & $15$
\\
$\Lambda(1810)$
&$\frac{1}{2}^+$
&
         & $0.9$ 
         & $\approx 0$ 
         & $2.8$ 
         & $6.3$
         & $0.7$ 
         & $0.7$ 
         & $3.9$ 
         & $2.3$
         & $1.7$
\\
$\Lambda(1830)$
&$\frac{5}{2}^-$
&
         & $0.6$ 
         & $2.2$ 
         & $0.4$ 
         & $1.6$
         & $2.0$ 
         & $1.8$ 
         & $1.6$ 
         & $1.3$
         & $5.3$
\\
\hline
\multicolumn{1}{c}{\small $\rightarrow \Sigma\eta $}&
\multicolumn{6}{l}{}\\
\hline
$\Sigma(1750)^1$
&$\frac{1}{2}^-$
&$\left(31.5\pm18.0\right)_{-4.5}^{+38.5}$ 
         & $-$ 
         & $-$ 
         & $-$ 
         & $-$
         & $5$ 
         & $11$ 
         & $25$ 
         & $71$
         & $$
\\
$\Sigma(1750)^2$
&$\frac{1}{2}^-$
&$\left(31.5\pm18.0\right)_{-4.5}^{+38.5}$ 
         & $3.0$ 
         & $3.1$ 
         & $1.5$ 
         & $5.0$ 
         & $0.6$
         & $3.8$ 
         & $4.7$ 
         & $1.9$
         & $$
\\
$\Sigma(1750)^3$
&$\frac{1}{2}^-$
&$\left(31.5\pm18.0\right)_{-4.5}^{+38.5}$ 
         & $6.0$ 
         & $2.1$ 
         & $25$ 
         & $10$
         & $3.8$ 
         & $1.4$ 
         & $14$ 
         & $6.3$
         & $3$
\\
$\Sigma(1775)$
&$\frac{5}{2}^-$
&
         & $\approx 0$ 
         & $0.05$ 
         & $\approx 0$ 
         & $\approx 0$
         & $0.02$ 
         & $0.01$ 
         & $\approx 0$ 
         & $\approx 0$
         & $$
\\
$\Sigma(1940)$
&$\frac{3}{2}^-$
&
         & $\approx 0$ 
         & $\approx 0$ 
         & $\approx 0$ 
         & $\approx 0$
         & $\approx 0$ 
         & $0.02$ 
         & $0.07$ 
         & $0.01$
         & $$
\\
\end{tabular}}
\end{ruledtabular}
\end{table*}
\renewcommand{\arraystretch}{1.0}

The nonrelativistic reduction again has a considerable effect on the $\eta$
decay widths. As observed in case of the $\Pi$ decays, it enhances in particular
the results for the $\frac{1}{2}^-$ states. The corresponding figures appear way
too big at least for the $\Lambda$(1670) and $\Lambda$(1800) resonances. On the
other hand, the decay widths of the $\frac{3}{2}^-$ are again reduced and
practically vanish. 

\section{Conclusions
\label{sec:summary}}

We have reported relativistic calculations of nonstrange decays of strange 
baryon resonances within CQMs. In particular, we have presented covariant
predictions for $\pi$ and $\eta$ decays of $\Lambda$, $\Sigma$, and $\Xi$
resonances by two types of CQMs, the ones with GBE and OGE dynamics.
The transition elements have been calculated
with a spectator-model decay operator in point-form relativistic quantum
mechanics. The present results complement the ones obtained earlier for
the same mesonic decay modes of the (nonstrange) $N$ and $\Delta$
resonances~\cite{Melde:2005hy}.

Regarding the $\pi$ decay widths we have found that the direct relativistic
predictions of the CQMs in general underestimate the experimental data. In
this respect the results parallel the ones obtained earlier for $N$ and
$\Delta$ resonances along the same approach. Here, only the $\Lambda$(1670)
represents an exception. We argue that a possible mixing effect with the
$\Lambda$(1405) resonance is responsible for this result. Such mixings of
resonance states should certainly be taken into account in a future more
refined calculation.

The systematics of the relativistic results has also led to a new
interpretation of the three lowest $\frac{1}{2}^-$ $\Sigma$ excitations
produced by the CQMs. In principle, all three can be seen as candidates
for the phenomenological $\Sigma$(1750). However, most naturally
only the third state $\Sigma(1750)^3$ is identified with the experimentally
measured $\Sigma$(1750), as it produces the most adequate $\pi$ decay width.

For the $\eta$ decay widths we have found a qualitatively similar behavior,
namely, they are all rather small or at most reach the (scarce)
experimental data from below. Even the largest $\eta$ decay width of
$\Lambda(1800)$ can be characterized in this manner, since its total
decay width is reported to be extremely large.

In the present work we have also shown that the PFSM decay operator has
a sensible nonrelativistic reduction, leading to the standard
elementary-emission model. However, it has also become evident that the
nonrelativistic decay widths exhibit no consistent pattern, as they vary
considerably in their magnitudes, scattering above and below the experimental
data. The nonrelativistic reduction introduces sizable effects strongly
depending on the $J^P$ values of the pertinent resonances. 

We have herewith completed the relativistic studies of $\pi$ and $\eta$
decays of light and strange baryon resonances within CQMs. We have established
the direct predictions of two types of CQMs without introducing any
additional parameterisations. The results show a consistent behavior but
they are not able to explain the experimental data. In any case relativistic
effects are found to be of utmost importance. The present study provides
for a reliable starting point to improve the relativistic description of
mesonic decays.

\begin{acknowledgments}
This work was supported by the Austrian Science Fund (FWF-Projects P16945
and P19035).
B. S. acknowledges support by the Doctoral Program 'Hadrons in Vacuum,
Nuclei and Stars' (FWF-Project W1203). The authors have profited from valuable 
discussions with L.~Canton, A.~Krassnigg, and R.F.~Wagenbrunn.
\end{acknowledgments}

\appendix
\section{Details of the nonrelativistic reduction}
{\label{sec:appendix}}

In the following we specify the nonrelativistic reduction of the PFSM calculation
that leads to the EEM. We leave the invariant phase-space factor in
Eq.~(\ref{eq:decwidth}) untouched and start with the matrix element of the
reduced decay operator in Eq.~(\ref{eq:transel})
\begin{multline}
\label{eq:a1}
F_{i\to f}=\\
\langle V',M',J',M_{J'},T',M_{T'}|{\hat D}^m_{rd}
|V,M,J,M_J,T,M_T \rangle =\\
\langle P',J',M_{J'},T',M_{T'}|{\hat D}^m_{rd}
|P,J,M_J,T,M_T \rangle \, ,
\end{multline}
which is now expressed in terms of momentum eigenstates $|P,J,M_J,T,M_T \rangle$.
In a first step we replace in this matrix element the Lorentz boosts by Galilean
boosts and use free three-quark
states $\left|{\vec{k}}_2,{\vec{k}}_3,{\vec{P}};\mu_1,\mu_2,\mu_3\right\rangle$
instead of velocity states for the representation of the eigenstates of
the quark-model Hamiltonian. This leads to baryon wave functions in the form
\begin{equation}
\label{eq:a2}
\langle {\vec k}'_2,{\vec k}'_3,{\vec P}';\mu'_1,\mu'_2, \mu'_3
|{\vec P},J,M_{J},T,M_{T}\rangle
\vspace{-0.1cm}
\nonumber
\end{equation}
\begin{equation}
   \qquad=
   \Psi_{MJ M_{J}TM_{T}}\left({\vec k}'_1,{\vec k}'_2,{\vec k}'_3;
\mu'_1,\mu'_2,\mu'_3\right)
     \delta^3\left({\vec P}'-{\vec P}\right) \, ,
\vspace{0.2cm}
\end{equation}
where the completeness relation of the free three-quark states reads
\begin{eqnarray}
\label{eq:a3}
{\mathbf 1}&=&\sum_{\mu_1,\mu_2,\mu_3}\int{d^{3}k_2d^{3}k_{3}d^3P}
\\
&&\times
\left|{\vec{k}}_2,{\vec{k}}_3,{\vec{P}};\mu_1,\mu_2,\mu_3\right\rangle
\left\langle  {\vec{k}}_2,{\vec{k}}_3,{\vec{P}};\mu_1,\mu_2,\mu_3\right|  
\nonumber  .
\end{eqnarray}
Using the latter one obtains for the spectator-model decay operator of 
Eq.~(\ref{momrepresent}) the following expression
\begin{widetext}
\begin{eqnarray}
&&
    {F_{i\to f}^{NR}}
    ={\langle P',J',M_{J'},T',M_{T'}|{\hat D}_{rd}^{m,NR}|
    P,J,M_J,T,M_T\rangle} 
\nonumber\\
&&
     {=\sqrt{2E}\sqrt{2E'}\sum_{\mu_i\mu'_i}
     \int{d^3k_2d^3k_3} 
     \Psi^\star_{M'J'M_{J'}T'M_{T'}}\left({\vec k}'_1,{\vec k}'_2,{\vec k}'_3;
     \mu'_1,\mu'_2,\mu'_3\right)    
} 
\nonumber\\
&&
     {\times  \frac{-3{\mathcal{N}}}{\sqrt{2p_{10}}\sqrt{2p'_{10}}} 
\frac{ g_{qqm}}{2m_1}\frac{1}{\sqrt{2\pi}}
\bar{u}(p_1',\mu_1')\gamma_5\gamma^\mu \mathcal{F}^m 
 u(p_1,\mu_1)q_\mu\delta_{\mu_2 \mu'_2}\delta_{\mu_3 \mu'_3}}
 \Psi_{MJM_JTM_T}\left({\vec k}_1,{\vec k}_2,{\vec k}_3;\mu_1,\mu_2,\mu_3\right)  ,
\label{eq:a4}
\end{eqnarray}
\end{widetext}
where $E$, $E'$  are the energies of the decaying and 
final baryons. Similarly, $p_{10}$ and $p'_{10}$ denote the energies
of the active quark in the incoming and outgoing channels, respectively.
The nonrelativistic baryon momenta satisfy
${\vec{P}}=\sum{\vec{p}}_i$ as well as
${\vec{P'}}=\sum{\vec{p}}'_i$. In addition, the energy and the
momentum of the emitted meson are given by $\omega_m=E-E'=p_{10}-p'_{10}$ and
${\vec{q}}={\vec{P}}-{\vec{P'}}={\vec{p}}_{1}-{\vec{p}}'_{1}$, respectively.

Next we have to express the various variables in Eq.~(\ref{eq:a4}) in terms
of the residual integration variables $\vec{k}_{2}$ and $\vec{k}_{3}$.
The corresponding relations are obtained
from a nonrelativistic limit of the original Lorentz boosts. This calculation
is conveniently carried out in the rest frame of the decaying baryon resonance
and leads to the following result
\begin{eqnarray}
&&{\vec p}_1=-{\vec k}_2-{\vec k}_3,
\nonumber\\
&&{\vec p}'_1=-{\vec k}_2-{\vec k}_3-{\vec q},
\nonumber\\
&&{\vec p}_2={\vec p}'_2={\vec k}_2,\nonumber\\
&&{\vec p}_3={\vec p}'_3={\vec k}_3,
\nonumber\\
&&{\vec k}'_1=-{\vec k}_2-{\vec k}_3-\frac{m_2+m_3}{ m_1+m_2+m_3}{\vec q},
\nonumber\\
&&
{\vec k}'_2={\vec k}_2+\frac{m_2}{m_1+m_2+m_3}{\vec q},\nonumber\\
&&
{\vec k}'_3={\vec k}_3+\frac{m_3}{m_1+m_2+m_3}{\vec q}.
\label{eq:a5}
\end{eqnarray}
Here, one has made the approximations $M \approx \sum m_i$
as well as $M' \approx \sum m'_i$, i.e. the interacting masses of the baryons
become equal to the free masses in the nonrelativistic limit; for the decay
modes considered in this paper one has furthermore $m_i=m'_i$. Furthermore,
one neglected terms of the orders
$(\frac{p_q}{m_q})^2$ and $(\frac{\omega_m}{m_q})^2$, upon the assumption 
that the quark masses $m_i$ are large as compared 
to the absolute value of the three-momentum ${\vec{p}}_i$ and the meson energy
$\omega_m$. 

For the practical calculation one transforms to a coordinate system, where
the momentum transfer to the final baryon is into the negative z-direction
and obtains for the relations of the primed  
and unprimed variables (${\vec{k}}'_2$, ${\vec{k}}'_3$) and 
(${\vec{k}}_2$, ${\vec{k}}_3$)
\begin{eqnarray}
\nonumber
k'_{ix}&=&k_{ix}\\
\nonumber
k'_{iy}&=&k_{iy}\\
k'_{iz}&=&k_{iz}+\frac{m_i}{m'_1+m_2+m_3}Q
\end{eqnarray}
for $i=2,3$. Here, $Q$ is the absolute value of the momentum transfer
$\vec q=(0,0,Q)$.
As a consequence of the reduction we also find $\omega_i \approx m_i$ 
and $\omega'_i \approx m'_i$, which reduces 
the normalisation factor $\mathcal N$ to 1. 
The final nonrelativistic reduction for the transition amplitude then reads
\begin{widetext}
\begin{eqnarray}
\label{NRdecwi}
&&
    {F_{i\to f}^{NR}}= {\sqrt{2E}\sqrt{2E'}\sum_{\mu_i\mu'_i}
     \int{d^3{\vec k}_2d^3{\vec k}_3} 
     \Psi^\star_{M'J'M_{J'}T'M_{T'}}\left({\vec k}'_1,{\vec k}'_2,{\vec k}'_3;
     \mu'_1,\mu'_2,\mu'_3\right)   
} 
\nonumber\\
&&
     {\times  \frac{-3 g_{qqm}}{2m_1} 
\frac{1}{\sqrt{2\pi}} \mathcal{F}^m 
\left\{
{\vec \sigma}_1
 \cdot {\vec{q}}-\frac{\omega_m}{2m_1}
{\vec \sigma}_1\cdot\left({\vec {p}}_1+{\vec {p}}'_1\right)
\right\}
 \delta_{\mu_2 \mu'_2}\delta_{\mu_3 \mu'_3}}
 \Psi_{MJM_JTM_T}\left({\vec k}_1,{\vec k}_2,{\vec k}_3;\mu_1,\mu_2,\mu_3\right)  \,  .
\end{eqnarray}
\end{widetext}
This result represents the familiar expression for the transition amplitude
in the EEM, where the terms proportional to 
$\vec \sigma_1\cdot{\vec{q}}$ and 
$\vec \sigma_1\cdot({\vec{p}}_1+{\vec{p}}'_1)$, 
involving the Pauli spin operator $\vec \sigma_1$ of the active quark,
are called direct and recoil terms, respectively 
(see, e.g., Refs.~\cite{LeYaouanc:1988aa,Ericson:1988gk}).
The latter is specific for the pseudovector coupling in the decay
operator~(\ref{momrepresent}); it would not be present at the same 
order of nonrelativistic approximation in case of pseudoscalar coupling. 

\addcontentsline{toc}{chapter}{Bibliography}

\end{document}